\documentclass[12pt]{article}

\usepackage[height=8.85in,width=6.45in]{geometry}
\usepackage{amsmath,amssymb}
\usepackage{hyperref}

\usepackage[utf8]{inputenc}
\usepackage{times}
\usepackage{courier}
\usepackage{braket}
\usepackage{mathtools}
\numberwithin{equation}{section}
\usepackage[svgnames]{xcolor}

\newcommand{\bR}{\mathbb{R}}
\newcommand{\cL}{\mathcal{L}}

\newcommand{\cO}{\mathcal{O}}
\newcommand{\vev}[1]{\langle#1\rangle}

\def\ren{\text{ren}}
\def\fC{\mathfrak{C}}
\def\fF{\mathfrak{F}}
\def\SO{\mathrm{SO}}

\usepackage{framed}
\definecolor{shadecolor}{rgb}{0.90,0.90,0.90}

\def\beq{\begin{eqnarray}}\def\eeq{\end{eqnarray}}
\def\be{\begin{equation}}\def\ee{\end{equation}}
\def\g{\gamma}
\def\r{\rho}
\def\s{\sigma}

\def\a{\alpha}
\def\e{\epsilon}

\def\b{\beta}
\def\d{\delta}

\def\D{\Delta}
\def\G{\Gamma}
\def\l{\lambda}

\def\ta{\tau}

\def\la{\langle}
\def\ra{\rangle}

\def\mo{{\mathcal{O}}}

\def\G{\Gamma}
\def\mc{{\mathcal{C}}}

\usepackage{bm}
\def\full#1{{
\text{\boldmath$#1$}}}

\begin{document}

\begin{titlepage}

\begin{flushright}
IPMU-17-0161
\end{flushright}

\vskip 4cm

\begin{center}

{\Large\bfseries First-order conformal perturbation theory \\[1em]
by marginal operators}

\vskip 1cm
Kallol Sen and Yuji Tachikawa
\vskip 1cm

\begin{tabular}{ll}
  & Kavli Institute for the Physics and Mathematics of the Universe, \\
& University of Tokyo,  Kashiwa, Chiba 277-8583, Japan\\
\end{tabular}

\vskip 1cm

\end{center}

\noindent 
We perform  conformal perturbation theory by marginal operators to first order. A suitable renormalization method is needed that makes the conformal invariance of the deformed correlation functions manifest. Combining the embedding space formalism with the dimensional regularization, we explicitly check that the deformed and renormalized two and three point functions are conformally invariant. 

\end{titlepage}

\tableofcontents

\section{Introduction}

Suppose that we are given a conformal field theory in $d$ dimensions, 
in the sense that the set of primary operators $\cO_i$ and their $n$-point functions \begin{equation}
\vev{\cO_{i_1}(x_1)\cdots \cO_{i_n}(x_n)}
\end{equation}
on $\bR^d$ are all specified, satisfying the usual consistency conditions.
 We do not necessarily have a Lagrangian $\cL$ for this theory, but we can still consider deforming the theory by adding to the would-be Lagrangian a small perturbation $\lambda \cO$ of scaling dimension $\Delta$:\footnote{%
 Here and in the following, we use the convention that when something is denoted by $A$ in the original theory, it is denoted in  bold as $\full{A}$ in the deformed theory, and we reserve the symbol $\delta$ to denote the order $\lambda$ part: 
$\full{A}=A+\lambda\delta A+O(\lambda^2)$.
 }\footnote{For an explicit application in the case of the AdS/CFT correspondence, see e.g.~\cite{Berenstein:2014cia},\cite{Berenstein:2016avf}.}
 \begin{equation}
 \cL\to \full{\cL}:=\cL+\lambda\cO . \label{wouldbe}
\end{equation}

By the deformation \eqref{wouldbe} of a would-be Lagrangian, we mean that we define the new un-renormalized correlation function by:\begin{equation}
\full{\vev{\cO_{i_1}(x_1)\cdots\cO_{i_n}(x_n)} }
:= \vev{\cO_{i_1}(x_1)\cdots \cO_{i_n}(x_n)} +
\l\int d^d x \vev{\cO(x)\cO_{i_1}(x_1)\cdots \cO_{i_n}(x_n)},
\end{equation}
where the integral is regularized  by a parameter $\epsilon$ in some specific way to deal with the divergence
when the position $x$ of the deforming operator approaches any of $x_i$ where other operators are inserted. 
This statement is explicitly verified for the case of two and three point functions in the deformed theory. 

We then introduce a renormalization \begin{equation}
\full{\cO_i^\ren} = Z^{-1}{}_i^j \full{\cO_j}
\end{equation} so that the deformed, renormalized $n$-point function \begin{equation}
\full{\vev{\cO_{i_1}^\ren (x_1)\cdots\cO_{i_n}^\ren(x_n)}}
\end{equation} is finite in the limit $\epsilon\to 0$ where the regularization is removed.

Conformal perturbation theory is about making this procedure precise. 
One common regularization is to remove a sphere of radius $\epsilon$ around each insertion $x_i$ in the integral, and then take the limit $\epsilon\to 0$. 
This regularization breaks conformal invariance, but preserves the Lorentz invariance. 
Therefore this method is enough for perturbations by relevant operators  for which we do not expect conformal invariance after the deformation.

Our aim in this paper is to perform the conformal perturbation by marginal operators, i.e.~the scalar operators $\cO$ of scaling dimension $\Delta=d$, up to first order in the perturbation parameter $\lambda$.
It is generally believed in the literature that such deformations will keep the conformal invariance.
To the authors' knowledge, this fact has not been explicitly demonstrated  for $d>2$ in the existing literature.
\footnote{For an analysis in $d=2$, see e.g.~\cite{Gaberdiel:2008fn}. 
With supersymmetry, there are various indirect arguments applying not to first order but to all orders.
There are also many works where the geometry of the space of exactly marginal couplings of supersymmetric theory is studied, but they are too numerous to be cited here.
Without supersymmetry, there are three recent works 
\cite{Bashmakov:2017rko,Behan:2017mwi,Hollands:2017chb}
where related questions were studied.
However, as far as the authors of this present paper can see, they did not actually show that the bootstrap equation is satisfied in their regularization scheme.
}

\goodbreak

In this short note, we concretely perform the following tasks:
\begin{itemize}
\item The first step is to find a regularization method which makes the conformal invariance of the  deformed renormalized $n$-point functions almost manifest.
This is done by combining the embedding space formalism and the dimensional regularization.
\item In the second step, we explicitly compute the deformed and renormalized two- and three-point functions, and confirm that they have the expected form dictated by the conformal symmetry.
\end{itemize}
We carry out this procedure in the simplifying situation when all the external operators are scalar.

The rest of the paper is organized as follows.
In Sec.~\ref{sec:pre}, we recall the embedding space formalism and the basic features of the operator product expansions (OPEs), mainly to set up our notations.
In Sec.~\ref{sec:reg}, we introduce our regularization method based on the analytic continuation in $\epsilon=\Delta-d$, show how a general $n$-point function is renormalized, and explicitly  work out the two-point and three-point functions.
We conclude the paper with various discussions in Sec.~\ref{sec:discussions}.

\section{Preliminaries}
\label{sec:pre}
We begin by recalling basic properties of $n$-point functions in any conformal field theory in $d$ dimensions. 
We work in a Wick-rotated Euclidean spacetime.

\subsection{Embedding space formalism}

To write down  $n$-point functions, we use the embedding space formalism.
For details, we refer the readers to the paper \cite{Costa:2011mg}.
Here we only recall what we will need in this paper.
We first promote   the position $x_\mu$ in $\bR^{d}$  to $X_M$ in $\bR^{d+1,1}$ with the metric \begin{equation}
X^2:=-X^+X^- + X^\mu X_\mu.
\end{equation}
We then impose $X^2=0$ and identify $X\sim c X$. 
The gauge freedom by this constant multiplication by $c$ can be fixed, if needed, by demanding $X^+=1$.

We then denote an operator $\cO_i$ at the position $X$ in the embedding space by $\cO_i(X)$.
We denote its scaling dimension by $\Delta_i$.
Inside the correlation function, we require that \begin{equation}
\cO_i(cX)=c^{-\Delta_i}\cO_i(X).
\end{equation}

Each operator $\cO_i$ belongs to a representation of $\SO(d)$, 
which we assume in this paper to be the $\ell_i$-index symmetric traceless tensor (STT). 
We simply call $\ell_i$ the spin of the operator $\cO_i$.

A spin-$\ell$ operator  $\cO_{\mu_1\cdots \mu_\ell}$ can be embedded by promoting the Lorentz indices from $\mu_i$ to $M_i$ and defining \begin{equation}
\cO(X,Z) := \cO_{M_1\cdots M_\ell}(X) Z^{M_1} \cdots Z^{M_\ell}.\label{OZ}
\end{equation}
The traceless condition in this language becomes the invariance\begin{equation}
\cO(X,Z+cX)=\cO(X,Z).
\end{equation}

The conformal invariance of an $n$-point function of conformal primaries \begin{equation}
\vev{\cO_{1}(X_1,Z_1)\cdots\cO_{n}(X_n,Z_n)}
\end{equation} translates in the embedding space formalism to the facts that 
\begin{itemize}
\item it scales by $c^{-\Delta_i}$ under the change $X_i \to cX_i$, and
\item it can be written in an $\SO(d+1,1)$ covariant manner using $(X_i)_M$ and $(Z_i)_M$.
\end{itemize}

We define the integral over the embedding space for a function $f(X)$ such that $f(cX)=c^{-d}f(X)$
by \begin{equation}
\int D^dX f(X) := \frac{1}{\mathrm{Vol\,GL}(1,\bR)^+}\int_{X^++X^-\ge 0} d^{d+2} X \delta(X^2) f(X)\,.
\end{equation}
This integral reduces to an ordinary integral over $\bR^d$, as can be seen by taking the gauge choice $X^+=1$.
The merit of the expression above is that it is manifestly $\SO(d+1,1)$ invariant, as long as the integral does not have any divergence issues.
We will come back to the analysis of the divergence later.

\subsection{Two-point functions}
The form of the two-point functions are determined uniquely by the conformal symmetry.
For scalars, we normalize the two-point functions so that we have 
\begin{equation}
\vev{\cO_{i}(X_1)\cO_{j}(X_2)}= \frac{\delta_{ij}}{X_{12}^{\Delta_{i}}},
\end{equation}
where  \begin{equation}
X_{ij}:=(X_i-X_j)^2=-2X_i\cdot X_j.
\end{equation}
Similarly, for spin-$\ell$ operators, we have \begin{equation}
{\la\mo_i(X_1,Z_1)\mo_{j}(X_2,Z_2)\ra}=\d_{ij}\frac{H_{12}^{\ell_i}}{X_{12}^{\ta_i}}\,,
\end{equation} where \begin{equation}
\tau_i:=\Delta_i + \ell_i
\end{equation} and \begin{equation}
H_{ij}=-2[(X_i\cdot X_j)(Z_i\cdot Z_j) -  (X_i\cdot Z_j)(X_j\cdot Z_i)].
\end{equation}

\subsection{Operator Product Expansions}\label{DOPE}

Conformal field theory admits operator product expansions (OPEs).
For example, the OPE of two scalar operators $\mo_1(X_1)$ and $\mo_2(X_2)$ takes the form
\begin{align}\label{undefope}
\begin{split}
\mo_i(X_1)  \mo_j(X_2)=\sum_{\mo_\a }
C_{ij\a }
\fC_{\Delta_i,\Delta_j,\Delta_\alpha,\ell_\alpha}\mo_\a (X_2,Z_2)\,
\end{split}
\end{align}

Here, $C_{ij\a}$ is the OPE coefficient depending on the theory,
and $\fC_{\Delta_i,\Delta_j,\Delta_\alpha,\ell_\alpha}$ is a  differential operator determined by the conformal Ward identities.\footnote{Objects denoted by German letters are differential operators.}
For brevity, we write the differential operator simply by $\fC$.

We normalize the differential operator $\fC$ so that the three-point function is given by \begin{equation}
\la\mo_i(X_1)\mo_j(X_2)\mo_{k}(X_3)\ra 
=  \frac{ C_{ijk}}{X_{12}^{\frac{\D_{i}+\D_{j}-\D_k}{2}}X_{23}^{\frac{\D_j+\Delta_{k}-\Delta_{i}}{2}}X_{31}^\frac{\D_k+\Delta_{i}-\Delta_{j}}{2}}
\end{equation} in the case of three scalars, and \begin{equation}
\la\mo_i(X_1)\mo_j(X_2)\mo_{k}(X_3,Z_3)\ra=\frac{C_{ijk}V_{3,12}^{\ell_k}}{X_{12}^\frac{\D_i+\D_j-\D_k-\ell_k}{2}X_{23}^\frac{\D_j+\D_k+\ell_k-\D_i}{2}X_{31}^\frac{\D_k+\ell_k+\D_i-\D_2}{2}}
\end{equation}
where \begin{equation}
V_{i,jk}=\frac{(Z_i\cdot X_j)(X_i\cdot X_j)-(Z_i\cdot X_k)(X_i\cdot X_k)}{X_{jk}}.
\end{equation}

We will also often use a shorthand notation
\be\label{f12l}
\fF_{ij\a}=C_{ij\a}\fC
\ee
so that the OPE is simply \begin{equation}
\mo_i(X_1)\mo_j(X_2)=\sum_{\mo_\a} \fF_{ij\a} \mo_\a(X_2,Z_2).
\end{equation}

The OPE of a scalar and an STT operator $\mo_i$ of dimension and spin $(\D_i,\ell_i)$ has the form
\be\label{spinope}
\mo_i(X_1)  \mo_{j}(X_2,Z_2)=\sum_{\mo_\a}\sum_m C_{ij\a}^m\fC_{m}\mo_{\a}\,,
\ee 
where  $\mo_\a$ can be scalars, STTs and other mixed-symmetry objects. 
The sum over $m$ represents the multiple tensor structures associated with this OPE. 
When $\mo_\a$ is STT, we normalize $\fC_m$ so that the three-point function is given by 
\begin{equation}
\la \mo_i(X_1,Z_1)\mo_j(X_2,Z_2)\mo_k(X_3)\ra=\sum_{m=0}^{\min(\ell_i,\ell_j)}C_{ijk}^m \frac{V_{1,23}^{\ell_i-m}V_{2,31}^{\ell_j-m}H_{12}^m}{X_{12}^{\frac{1}{2}(\ta_1+\ta_2-\ta_3)}X_{13}^{\frac{1}{2}(\ta_1+\ta_3-\ta_2)}X_{23}^{\frac{1}{2}(\ta_2+\ta_3-\ta_1)}}\,.
\end{equation}

\section{Regularization and renormalization}
\label{sec:reg}

In this section we set up the regularization and the renormalization of the $n$-point functions.
Our regularization method is the dimensional regularization,
where we make an analytic continuation of  the difference $\epsilon=\Delta-d$ away from zero.
We remind the reader that  $\Delta$ is the scaling dimension of the deformation operator $\cO$
and $d$ is the spacetime dimensionality.

\subsection{Two point functions}
Let us start by studying the deformation of the scalar two-point function.
The deformed unrenormalized two point functions are
\be
\full{\la\mo_{i}(X_1)\mo_{j}(X_2)\ra}=\la\mo_{i}(X_1)\mo_{j}(X_2)\ra+\l\int D^d X\la\mo(X)\mo_{i}(X_1)\mo_{j}(X_2)\ra\,
\ee
where the three point function is given by 
\begin{align}
\la\mo(X)\mo_{i}(X_1)\mo_{j}(X_2)\ra 
=&  \frac{ C_{\mo ij}}{X_{12}^{\frac{\D_{i}+\D_{j}-\D}{2}}X_{01}^{\frac{\D+\Delta_{j}-\Delta_{i}}{2}}X_{02}^\frac{\D+\Delta_{i}-\Delta_{j}}{2}}\,.
\end{align} 
Here and below, $X_{0i}=(X-X_i)^2=-2X\cdot X_i$.

The integral over $X$ coordinate with an analytic continuation $\Delta-d=\epsilon$
can be explicitly evaluated, see Appendix~\ref{2ptfn}.
We find that it is zero when $\Delta_{i}\neq \Delta_{j}$,
while 
if $\Delta_{i}=\Delta_{j}$ we have
\begin{align}
\int D^d X\la\mo(X)\mo_{i}(X_1)\mo_{j}(X_2)\ra
& =  S_{d-1} C_{\mo{i}{j}}\frac{2}{d-\D}\frac{1}{X_{12}^{\D_{i}+\frac{\D-d}{2}}} +O(\epsilon) \\
&= \frac{\mc_{\mo ij}}{X_{12}^{\D_{i}}}\bigg(-\frac{2}{\e}+\log X_{12}\bigg)+O(\epsilon)\,,
\label{bosh}
\end{align}
where 
\be\label{normsc}
\mc_{\mo{i}{j}} := S_{d-1} C_{{i}{j}\mo}.
\ee
with $S_{d-1}$ the volume of the unit sphere $S^{d-1}$.

To simplify the presentation below,
we assume that $C_{\mo ij}$ is diagonalized as a matrix  whose indices are $i$ and $j$,
by a suitable choice of basis.
We then define the renormalized operators 
\begin{equation}
\full{\mo_i^\ren} :=
Z_i^{-1}\mo_i,\qquad
Z_i^{-1}=(1+\lambda\mc_{ii\mo} /\epsilon) 
\label{Z}
\end{equation}
Then we have 
\begin{align}
\full{\vev{ \mo_{i}^\ren(X_1)\mo_{j}^\ren(X_2)}}
=\frac{\delta_{ij}}{X_{12}^{\full{\Delta_{i}}}}\,,
\end{align} 
where 
\begin{equation}
\full{\Delta_{i}}= \Delta_{i} - \lambda\mc_{\mo ii}\,.\label{D}
\end{equation}

The computations for the spin-$\ell$ case can be carried out in a similar manner, see Appendix \ref{STT2}. 
We end up having the same formula for the wavefunction renormalization factor $Z_{i}^{-1}$ 
and the deformed scaling dimension \eqref{D} except that now $\mc_{\mo ij}$ is given by
\begin{equation}
\mc_{\mo ij}=\G(\frac d2) S_{d-1}\sum_{m=0}^{\ell_\rho}C_{\mo ij}^m \frac{\G(\ell_i-m+1)}{\G(\frac d2+\ell_i-m)}\,.
\end{equation}
We again assume below that a suitable basis is chosen so that $\mc_{\mo ij}$ is diagonal as a matrix whose indices are $i$ and $j$.

\subsection{General $n$-point functions}\label{DCORR}
Let us now discuss the regularization and the renormalization of a general $n$-point function.
We  need to analyze the divergent part of the integral over $X$ of $\la\mo(X) \mo_{i_1}\cdots\mo_{i_n}\ra$. 
The divergence arises from the coincident limits of the deformation operator and each of the external operators.
These limits can be studied using OPE.
For simplicity, we present the discussion assuming that all $\cO_{i_j}$ are scalar.

When $\cO(X)$ hits $\cO_{i_1}(X_1)$, we have
\begin{equation}
\la\mo(X) \mo_{i_1}\cdots\mo_{i_n}\ra
=\sum_{\mo_\a}\fF_{\mo i_1\a}\la\mo_\a(X_1,Z_1)\mo_{i_2}\cdots\mo_{i_n}\ra\,.
\label{decomp}
\end{equation}
For the moment we analyze the divergence in the original physical space.
An $n$-th descendant of the operator $\cO_\a$ has a divergence of the form 
\begin{equation}
\sim 1/{r^{\D+\D_{i_1}-(\D_\a-\ell_\a+n)}}\,,\label{rr}
\end{equation} where $r=|x-x_{i_1}|$.
As $r\rightarrow0$,  the integral of \eqref{rr} diverges when 
\begin{equation}
\D+\D_{i_1}-(\D_\a-\ell_\a+n) \ge d\,.
\end{equation}

Most of the divergence is rendered harmless by the dimensional regularization, using an analytic continuation \begin{equation}
\Delta-d = \epsilon \neq 0\,,
\end{equation} and finally taking $\epsilon\to 0$. 
This procedure fails exactly when \begin{equation}
\D_{i_1}-(\D_\a-\ell_\a+n) = 0\,, 
\end{equation} 
in which case a $1/\epsilon$ divergence remains.
Below, we call such a divergence, scaling as $\sim 1/r^\Delta$ in the $r\to 0$ limit as a \emph{dangerous divergence}.
Other divergences and the finite pieces are called \emph{safe}.
To proceed further, we make an important genericity assumption in this paper:
\begin{shaded}
\noindent\textbf{Assumption:} \emph{The original theory is generic enough so that no two primary operator $\cO_{i}$ and $\cO_j$ with nonzero $C_{\cO ij}$ satisfies $\D_{i}-\ell_i = \D_{j}-\ell_j+n$ for any nonzero integer $n$. }
\end{shaded}
Then the only divergence in \eqref{decomp} which cannot be removed by the dimensional regularization is when the operator $\cO_\a$ appearing in the intermediate channel
is the primary $\cO_{i_1}$ itself. 
This contributes to the  divergence of the form \begin{equation}
\la\mo(X) \mo_{i_1}(X_1)\cdots\mo_{i_n}(X_n)\ra
\sim \frac{C_{\mo i_1i_1}}{r^{\Delta}} \la \mo_{i_1}(X_1)\cdots\mo_{i_n}(X_n)\ra.
\end{equation}
where the symbol $\sim$ means that the difference of the left hand side and the right hand side is less divergent. 

Then, the integral of $X$ close to $X_1$ gives a $1/\epsilon$ divergence in the deformed unrenormalized $n$-point function of size \begin{equation}
-\frac{\mc_{\mo i_1i_1}}{\epsilon} \la \mo_{i_1}(X_1)\cdots\mo_{i_n}(X_n)\ra
=\frac{\delta\Delta_i}{\epsilon} \la \mo_{i_1}(X_1)\cdots\mo_{i_n}(X_n)\ra
\end{equation}
In other words,
 \begin{equation}
\int D^d X\la\mo(X)\mo_{i_1}\cdots\mo_{i_n}\ra' := \int D^d X\la\mo(X)\mo_{i_1}\cdots\mo_{i_n}\ra-\frac{1}{\e}\sum_{j=1}^n \delta\Delta_{i_j} \la\mo_{i_1}\cdots\mo_{i_n}\ra\,,\label{fin}
\end{equation} is finite in the limit $\epsilon\to 0$.
We find that  the wavefunction renormalization introduced in \eqref{Z} removes the $1/\epsilon$ divergences properly and  that
\begin{align}
\begin{split}
\full{\vev{\mo_{i_1}^\text{ren}\cdots\mo_{i_n}^\text{ren}}}
=&\la\mo_{i_1}\cdots\mo_{i_n}\ra+\l \int D^d X\la\mo(X)\mo_{i_1}\cdots\mo_{i_n}\ra'\,.
\end{split}
\end{align}
In the next section, we carry out the explicit construction of the deformed and renormalized three point functions. To ensure that the deformed and renormalized four point functions, constructed using the formalism of this work, satisfy the bootstrap identities, is subtle and needs some thought.

\subsection{Three-point functions}\label{Threept}
Let us see the formalism just presented in action, in the case of the scalar-three point functions.
We first note that the dangerous divergence of the four-point function
\begin{equation}
\la\mo(X)\mo_{i}(X_1)\mo_{j}(X_2)\mo_{k}(X_3)\ra \label{4pt}
\end{equation} when $X\to X_1$ is the same as the dangerous divergence of the function
\begin{equation}
P_{123}(X,X_1,X_2,X_3):=
C_{\mo i i}\frac{ X_{12}^{\frac{\Delta}4}X_{13}^{\frac{\Delta}4}}{X_{01}^\frac{\D}{2} X_{02}^{\frac\Delta4}X_{03}^{\frac\Delta4}}
\frac{C_{ijk}}{X_{12}^\frac{\D_i+\D_j-\D_k}{2}X_{13}^\frac{\D_i+\D_k-\D_j}{2}X_{23}^\frac{\D_j+\D_k-\D_i}{2}} ,\label{boo}
\end{equation}
which is essentially the contribution to \eqref{4pt}
from the OPE of $\cO$ with $\cO_{i}$ with the same operator $\cO_{i}$ in the intermediate channel.
Note that the equation \eqref{boo} has the correct scaling properties to be thought of as a function on the embedding space.

We then decompose the original 4-point function as  \begin{multline}
\la\mo(X)\mo_{i}(X_1)\mo_{j}(X_2)\mo_{k}(X_3)\ra
=\\ 
\la\mo(X)\mo_{i}(X_1)\mo_{j}(X_2)\mo_{k}(X_3)\ra^\circ
+\sum_{(abc)=(123),(231),(312)} P_{abc}(X,X_a,X_b,X_c).
\end{multline}
The part $\la\mo(X)\mo_{i}(X_1)\mo_{j}(X_2)\mo_{k}(X_3)\ra^\circ$ has no dangerous divergence, and behaves correctly under the scaling $X\to cX$ and $X_i\to c_i X_i$.
Therefore, its integral over $X$ is guaranteed to be of the following form \begin{equation}
\int D^dX \la\mo(X)\mo_{i}(X_1)\mo_{j}(X_2)\mo_{k}(X_3)\ra^\circ
= \frac{c_{ij k}}{ X_{12}^\frac{\D_i+\D_j-\D_k}{2}X_{13}^\frac{\D_i+\D_k-\D_j}{2}X_{23}^\frac{\D_j+\D_k-\D_i}{2} } \label{xxx}
\end{equation}
where $c_{ijk}$ is a constant defined by this integral.

The integral of the functions $P_{abc}(X,X_a,X_b,X_c)$ can be computed explicitly using dimensional regularization using the formulas given in Appendix~\ref{sec:useful}. 
We find \begin{multline}
\int D^dX P_{123}(X,X_1,X_2,X_3)
= \\
\frac{\mc_{\mo ii} C_{ij k}}{ X_{12}^\frac{\D_i+\D_j-\D_k}{2}X_{13}^\frac{\D_i+\D_k-\D_j}{2}X_{23}^\frac{\D_j+\D_k-\D_i}{2} }
\left[
-\frac1\epsilon
+\frac{1}{2}\log \frac{X_{12}X_{13}}{X_{23}} + H_d+O(\epsilon).
\right]
\label{qqq}
\end{multline}
where the constant $H_d$ is given by
\be\label{Hfn}
H_d=-\g+\psi\bigg(\frac{d}{2}\bigg)-2\psi\bigg(\frac{d}{4}\bigg)\,
\ee where $\psi$ is the di-gamma function and $\g$ is the Euler-Mascheroni constant.

Combining all contributions, we see that 
\begin{equation}
\full{\vev{\mo_{i}^\ren\mo_{j}^\ren\mo_{k}^\ren}
= \frac{C_{ijk}}{
X_{12}^{\frac{\D_{i} +\D_{j} -\D_{k} }{2}}
X_{23}^{\frac{\D_{j} +\D_{k} -\D_{i} }{2}}
X_{31}^{\frac{\D_{k} +\D_{i} -\D_{j} }{2}}
}}
\end{equation}
where \begin{equation}
\full{C_{ijk}}=\bigg[1+\lambda(\mc_{\mo ii}+\mc_{\mo jj}+\mc_{\mo kk}) H_d\bigg]C_{ijk}+\lambda c_{ijk} + O(\lambda^2)\,
\end{equation}
where $c_{ijk}$ is defined in \eqref{xxx} and $H$ is given in \eqref{Hfn} .

This further implies that we have an equality \begin{equation}
\full{\vev{\mo_{i}^\ren\mo_{j}^\ren\mo_{k}^\ren}=\sum_{k'}\fF_{ijk'} \vev{\mo_{k'}^\ren\mo_{k}^\ren}},\label{deformed3ptOPE}
\end{equation} where \begin{equation}
\full{\fF_{ijk} =C_{ijk} \fC }
\end{equation}
i.e. at least within the deformed renormalized three-point function,
we can perform the OPE using the deformed parameters.
The analysis above can be extended to the case of the scalar-scalar-STT 3-point function in a straightforward manner. The details can be found in Appendix~\ref{STT}.

\section{Discussions}
\label{sec:discussions}

In this short note, we studied conformal perturbation theory of a given conformal field theory in $d$ dimension by a marginal operator $\cO$ to a linear order.
We  computed the deformed and renormalized 2-point and 3-point functions explicitly
and showed that they have the correct form expected from a conformal field theory.
We end this paper by listing below three natural directions of further research.

\subsection{Operators with more general $\SO(d)$ representations}
It would be nice to show the validity of the bootstrap equation for the external operators in a complete general representation of $\SO(d)$.
Our strategy of the proof was the following:
we first computed in Sec.~\ref{sec:reg} explicitly the deformed and renormalized two-point functions  for scalar and STT operators,
and the deformed and renormalized three point functions for the scalar-scalar-scalar case and the scalar-scalar-STT case.

For generalization, we just need to establish that the two-point functions and the three-point functions with operators of general $\SO(d)$ representation of the deformed theory have the correct form as dictated by the conformal symmetry.
In principle, this can be done by an explicit computation as was done in this paper,
but it might be nicer to have a more abstract argument guaranteeing the appearance of the correct form.
So far the authors could only show the appearance of the correct logarithms in \eqref{qqq} only by an explicit evaluation of the integral.

\subsection{Removing the genericity assumption}
In Sec.~\ref{DCORR}, to simplify the discussion of how the theory is renormalized,
we assumed a genericity condition that no two primary operators $\cO_i$ and $\cO_j$ have scaling dimensions separated by a nonzero integer.
It would be nice to remove or at least weaken this assumption.
It might also be possible that there are in fact a certain conformal field theory such that the perturbation by a marginal operator to linear order is already not a conformal field theory, contrary to the popular belief.
In either way, a further analysis in this direction would be interesting.

\subsection{The inverse problem}
In this paper, we demonstrated that the first order deformation by a marginal operator of a conformal field theory is still a conformal field theory, in the sense that the deformed and renormalized two and three point functions are conformally invariant. 
It would be interesting to ask the inverse problem. 

Namely, suppose a continuous one-parameter family $\mathrm{CFT}_\lambda$ of conformal field theory is given. Is this one-parameter family always given by a deformation by a marginal operator?
The answer to this question depends on what we mean by the phrase \emph{specifying a one-parameter family $\mathrm{CFT}_\lambda$ of conformal field theory}.
If we take this phrase to mean that \emph{we give functions $\Delta_i(\lambda)$ and $C_{ijk}^m(\lambda)$ satisfying the bootstrap equations for all $\lambda$},
the question is far from obvious, and would be indeed an interesting question in mathematical physics.

However, if we take the point of view that \emph{for any one-parameter family of a quantum field theory,
we should be able to vary the coupling constant (such as $\lambda$) spatially}, 
the inverse problem is almost tautologically true.
This is because, by definition, we can consider the situation such that the system is described by $\mathrm{CFT}_{\lambda=\lambda_0}$ around the origin of the spacetime
while at the asymptotic infinity the system is given by $\mathrm{CFT}_{\lambda=0}$.
Then, we can shrink the region where the parameter $\lambda$ is turned on by a scaling transformation.
By the state-operator correspondence of the conformal field theory, this would give the deforming operator $\cO$.

There is in fact an interesting counterexample to this inverse question, interpreted in the former manner.\footnote{The authors thank Nati Seiberg to point out this counter-example.}
Consider the quantum two-dimensional Liouville theory, which comes in a family parameterized by $b$, the strength of the exponential interaction $e^{-b\varphi}$.
The derivative of this interaction with respect to $b$ is not a sensible operator.
Therefore, this family is \emph{not} given by a deformation by an exactly marginal operator.
It is not clear to the authors at present what is the essential feature of this counter-example:
is it because that dimension 2 is special, or is it because of the noncompact nature of the theory?
It would be interesting to establish a sensible necessary condition so that the inverse problem can be answered positively.

\section*{Acknowledgments}
KS would like to thank the Yukawa Institute for Theoretical Physics
at Kyoto University for hospitality during the workshop YITP-W-17-08
"Strings and Fields 2017," for hospitality during the course of this work.
YT is partially supported in part byJSPS KAKENHI Grant-in-Aid (Wakate-A), No.17H04837 
and JSPS KAKENHI Grant-in-Aid (Kiban-S), No.16H06335.
KS and YT  are partially supported by WPI Initiative, MEXT, Japan at IPMU, the University of Tokyo.

\appendix

\section{Useful formulas}
\label{sec:useful}
\subsection{The master formula}
A large part of the calculation in this note entails performing the following type of integral,
\begin{equation}\label{int}
\int D^{d}X  \frac{1}{X_{01}^a X_{02}^b X_{03}^c}\,,
\end{equation}
and its variants. 
This type of integrals, called the conformal integrals, have been considered in detail in \cite{SimmonsDuffin:2012uy}. The idea is to use Feynman parameterization and then perform the $X$ integral with the generic form
\be\label{int3}
\int D^d X \frac{1}{(-2X\cdot Y)^\d}\,,
\ee
where the integrand itself is well defined only for $\d=d$. 
This implies that $a+b+c=2h$ should hold for \eqref{int1}, where we use the standard notation $h=d/2$.
For us, this is unfortunately not the case since $a+b+c=\D$ (the dimension of the deformation operator) and this is not $d$ once we take the analytic continuation. 
To make \eqref{int3} well defined when $\d\neq d$, we modify \eqref{int3} by a factor $(X^+)^{\d-d}$ so that under the gauge $X^+=1$,  
\be\label{intt}
\int D^d X\frac{(X^+)^{\d-d}}{(-2X\cdot Y)^\d}=S_{d-1}\frac{\G(h)\G(\d-h)}{2\G(\d)}\frac{(Y^+)^{\d-d}}{(-Y^2)^{\d-h}}\,,
\ee
where $S_{d-1}$ is the volume of the unit sphere $S^{d-1}$ and $h=d/2$ as always. When $\d=d$, the numerator on the \textit{rhs} of \eqref{intt} becomes 1. Using \eqref{intt}, we can convert \eqref{int} into,
\be\label{int1}
\int D^d X \frac{(X^+)^{a+b+c-2h}}{X_{01}^a X_{02}^b X_{03}^c}=S_{d-1}\frac{\G(a+b+c-h)\G(h)}{2\G(a)\G(b)\G(c)}\int_0^\infty \frac{d\a}{\a}\frac{d\b}{\b}\frac{\a^{b-1}\b^{c-1}(Y^+)^{a+b+c-2h}}{(\a X_{12}+\b X_{13}+\a\b X_{23})^{a+b+c-h}}\,,
\ee
where $Y^+=1+\a+\b$. For conformal integrals in \cite{SimmonsDuffin:2012uy}, this factor is not present since $a+b+c=d$. For our case, however, $a+b+c=\D\rightarrow d$. To compute the integral in \eqref{int1}, we write, 
\be
(Y^+)^{a+b+c-2h}=\sum_{m,n=0}^\infty\frac{\G(\D-2h+1)}{\G(\D-2h-m-n+1)m!n!}\a^m \b^n\,,
\ee
and perform the integrals over $\a$ and $\b$, to obtain, 
\begin{align}
\begin{split}
\int D^d X\frac{(X^+)^{a+b+c-2h}}{X_{01}^a X_{02}^bX_{03}^c}=&\frac{\G(h)S_{d-1}}{X_{12}^b X_{13}^c}\sum_{m,n=0}^\infty\frac{\G(\D-2h+1)}{\G(\D-2h-m-n+1)m!n!}\\
&\times \bigg(\frac{X_{23}}{X_{12}X_{13}}\bigg)^{a-h}\bigg(\frac{X_{12}^m X_{13}^n}{X_{23}^{m+n}}\bigg)\G(h-a+m+n)(b)_{a-h-m}(c)_{a-h-n}\,.
\end{split}
\end{align}
For our purposes, $a=\D/2$ and for a generic three point function, $c>0$. For $m+n\geq1$, all the terms in the above sum are $O(\e)$ suppressed and we have a non-zero contribution only from $m=n=0$, so that finally,
\be\label{int4}
\int D^d X\frac{(X^+)^{a+b+c-2h}}{X_{01}^a X_{02}^bX_{03}^c}=\G(h)S_{d-1}\frac{1}{X_{12}^b X_{13}^c}\bigg(\frac{X_{23}}{X_{12}X_{13}}\bigg)^{a-h}\frac{\G(a+b-h)\G(h-a)\G(a+c-h)}{2\G(a)\G(b)\G(c)}\,.
\ee
When $c=0$, which is the case for the integral over the three point functions (or equivalently for the deformed two point functions), we can take either \eqref{int} and set $c=0$ from the start, or consider \eqref{int4} and put $c=0$, $X_3=\infty$. A slight variant of \eqref{int}, useful for the purpose of dealing with external spin operators, is given by 
\begin{align}
\begin{split}
&\int D^d X (X^+)^{a+b+c-2h}\frac{V_{3,1p}^f V_{3,2p}^g}{X_{1p}^a X_{2p}^b X_{3p}^c}\\
&=\frac{S_{d-1}\G(h)\G(a+b+c-h)}{\G(a+f)\G(b+g)\G(c)\G(-f)\G(-g)}\int \frac{d\a_1}{\a_1}\frac{d\a_2}{\a_2}\frac{d\a_4}{\a_4}\frac{d\a_5}{\a_5}\a_1^{a+f}\a_2^{b+g}\a_4^{-f}\a_5^{-g}\\
&\times\frac{(Y^+)^{a+b+c-2h}}{(\a_1 X_{13}+\a_2X_{23}+\a_1\a_2 X_{12}+(\a_2\a_4-\a_1\a_5) V_{3,12}X_{12})^{(a+b+c-h)}}\,,
\end{split}
\end{align}
where $Y^+=1+\sum_{i=1}^3 \a_i+\a_4 C_{Z_3,X_1,X_3}^++\a_5 C_{Z_3,X_2,X_3}^+$. We can decompose $(Y^+)^{a+b+c-2h}$ in terms of the sum as in the case of \eqref{int1} but only the first term in the sum eventually contributes since the other terms are $O(\e)$ suppressed. Hence,

\begin{align}\label{intspin}
\begin{split}
&\int D^d X \frac{V_{3,1p}^f V_{3,2p}^g}{X_{1p}^a X_{2p}^b X_{3p}^c}\\
&=(-1)^g S_{d-1} X_{12}^{c-h}X_{13}^{-a-c+h}X_{23}^{-b-c+h}V_{3,12}^{f+g}\frac{\G(h)\G(a+c-h)\G(b+c-h)\G(f+g+h-c)}{2\G(c)\G(a+f)\G(b+g)}\,.
\end{split}
\end{align} 

\subsection{Example: scalar two-point functions}\label{2ptfn}
Let us consider the specific case of the deformed two point functions or equivalently the integral over the three point function.
We need to perform the integral over the three point function:
\begin{align}\label{scal2pt}
\begin{split}
\int D^d X \la\mo_{i}(X_1)\mo_{j}(X_2)\mo(X)\ra&=\frac{C_{ij\mo}}{X_{12}^\frac{\D_i +\D_j-\D}{2}}\int D^d X\frac{1}{X_{10}^\frac{\D+\D_{ij}}{2}X_{20}^\frac{\D-\D_{ij}}{2}}\\
&=\frac{C_{ij\mo}}{X_{12}^\frac{\D_i +\D_j+\D-2h}{2}}\frac{S_{d-1}\G(h)\G(\D-h)}{2\G(\frac{\D+\D_{ij}}{2})\G(\frac{\D-\D_{ij}}{2})}\int_0^\infty \frac{d\a}{\a}\a^\frac{d-\D-\D_{ij}}{2}(1+\a)^{\D-d}\\
&=\frac{C_{ij\mo}}{X_{12}^\frac{\D_i +\D_j+\D-2h}{2}}\frac{S_{d-1}\G(h)\G(\D-h)}{2\G(\frac{\D+\D_{ij}}{2})\G(\frac{\D-\D_{ij}}{2})}\frac{\G(\frac{2h-\D-\D_{ij}}{2})\G(\frac{2h-\D+\D_{ij}}{2})}{\G(2h-\D)}\,.
\end{split}
\end{align}
Notice that whenever $\D_{ij}:=\D_i-\D_j\neq0$, \eqref{scal2pt} is $O(\e)$ suppressed. When $\D_{ij}=0$, we have,
\be\label{eqsc}
\int D^d X \la\mo_{i}(X_1)\mo_{j}(X_2)\mo(X)\ra=\frac{C_{ii\mo}}{X_{12}^{\D_i}}S_{d-1}\bigg(-\frac{2}{\e}+\log X_{12}\bigg)+O(\e)\,.
\ee
which gives \eqref{bosh}.

\section{Deformed two and three point functions for STT operators}
First we consider the deformed two point functions for external STT operators of spin $\ell$ and next the deformed three point functions involving two scalars and an STT operator. 

\subsection{Deformed two point functions}\label{STT2}

The argument for the deformed two point functions for the external spin operators will be the same as for the scalar case. The only difference will be the additional tensor structures. We will consider the integral,
\be
\int D^d X \la\mo_\rho(Z_1,X_1)\mo_{\rho'}(Z_2,X_2)\mo(X)\ra=\sum_{m=0}^{\text{min}(\ell_\rho,\ell_{\rho'})} \frac{C_{\rho\rho'\mo}^m}{X_{12}^\frac{\ta_\rho+\ta_{\rho'}-\D}{2}} H_{12}^m  \int D^d X\frac{V_{1,20}^{\ell_\rho-m}V_{2,10}^{\ell_{\rho'}-m}}{X_{10}^\frac{\D+\ta_{\rho\rho'}}{2} X_{20}^\frac{\D-\ta_{\rho\rho'}}{2}}
\ee
where the dimensions and spin quantum numbers for $\mo_\rho$ is $(\D_\rho,\ell_\rho)$ and for $\mo_{\rho'}$ is $(\D_{\rho'},\ell_{\rho'})$, and $\ta_\rho=\D_\rho+\ell_\rho$ and $\ta_{\rho\rho'}=\ta_\rho-\ta_{\rho'}$. We can convert into the Feynman parameterization and perform some of the integrals so that\footnote{ In general $Y^+=1+\a+\b C_{Z_1,X_2,X_1}^++\g C_{Z_2,X_1,X_2}^+$. We are choosing a gauge such that $C_{Z_1,X_2,X_1}^+=0$ and $C_{Z_2,X_1,X_2}^+=1$ so that $\b$ drops out from $Y^+$. } ,
\begin{align}
\begin{split}
&\int D^d X \la\mo_\rho(Z_1,X_1)\mo_{\rho'}(Z_2,X_2)\mo(X)\ra\\
&=\frac{H_{12}^{\ell_\rho}}{X_{12}^\frac{\ta_\rho+\ta_{\rho'}+\D-d}{2}} \sum_{m=0}^{\text{min}(\ell_\rho,\ell_{\rho'})}C_{\rho\rho'\mo}^m \frac{(-1)^{\ell_\rho-m}\G(\D-h-m+\ell_\rho)}{2\G(m-\ell_{\rho'})\G(\ell_\rho-m+\frac{\D-\ta_{\rho\rho'}}{2})\G(\ell_{\rho'}-m+\frac{\D+\ta_{\rho\rho'}}{2})}\\
&\times\G(h) S_{d-1}\int d\a d\g\ \a^{\frac{d-\D-\ta_{\rho\rho'}}{2}-1}(1+\a+\g)^{\D-d}\g^{\ell_\rho-\ell_{\rho'}-1}\\
&=\frac{H_{12}^{\ell_\rho}}{X_{12}^\frac{\ta_\rho+\ta_{\rho'}+\D-d}{2}} \sum_{m=0}^{\text{min}(\ell_\rho,\ell_{\rho'})}C_{\rho\rho'\mo}^m\frac{\G(1+\ell_{\rho'}-m)\G(\D-h-m+\ell_\rho)}{\G(1+\ell_{\rho'}-\ell_\rho)\G(\ell_\rho-m+\frac{\D-\ta_{\rho\rho'}}{2})\G(\ell_{\rho'}-m+\frac{\D+\ta_{\rho\rho'}}{2})}\\
&\times \frac{\G(\frac{2h-\D-\ta_{\rho\rho'}}{2})\G(\frac{2h-\D+\s_{\rho\rho'}}{2})}{2\G(2h-\D)}\,.
\end{split}
\end{align}
where $\s_{\rho\rho'}=(\D_\rho-\ell_\rho)-(\D_{\rho'}-\ell_{\rho'})$. With the assumption that no two operators in the spectrum have integer separation of dimension, observe that only for $\s_{\rho\rho'}=0$ and $\ta_{\rho\rho'}=0$ contributes to the leading divergence and $O(\e)$ suppressed otherwise. So that,  
\begin{align}\label{spin2pt}
\begin{split}
&\int D^d X \la\mo_\rho(Z_1,X_1)\mo_{\rho}(Z_2,X_2)\mo(X)\ra\\
&=\frac{H_{12}^{\ell_\rho}}{X_{12}^{\ta_\rho+\frac{\D-d}{2}}}\frac{2\G(h)}{d-\D}S_{d-1} \sum_{m=0}^{\ell_\rho}C_{\rho\rho\mo}^m \frac{\G(\ell_\rho-m+1)\G(-h-m+\D+\ell_\rho)}{\G(\ell_\rho-m+\frac{\D}{2})^2}\\
&=\frac{H_{12}^{\ell_\rho}}{X_{12}^{\ta_\rho}}\mc_{\rho\rho'\mo}\bigg(-\frac{2}{\e}+\log X_{12}\bigg)+O(\e)\,,
\end{split}
\end{align}
where
\be
\mc_{\rho\rho'\mo}:=\G(h) S_{d-1}\sum_{m=0}^{\ell_\rho}C_{\rho\rho'\mo}^m \frac{\G(\ell_\rho-m+1)}{\G(h+\ell_\rho-m)}\,.
\ee
Similar to the scalar case, we assume that $\mc_{\rho\rho'\mo}$ is diagonalized as a matrix with indices $\rho$ and $\rho'$. 
We then  define the renormalized operators:
\be\label{Zspin}
\full{\mo_\rho}:=Z_\rho^{-1}\mo_\rho\,, \qquad Z_\rho^{-1}=(1+\l \mc_{\rho\rho\mo}/\e)\,,
\ee
so that 
\be
\full{\la\mo_\rho(Z_1,X_1)\mo_{\rho'}(Z_2,X_2)\ra}=\d_{\rho\rho'}\frac{H_{12}^{\ell_\rho}}{X_{12}^\full{\ta_\rho}}\,,
\ee
where
\be
\full{\ta_\rho}=\ta_\rho-\l \mc_{\rho\rho\mo}\,.
\ee
For $\ell_\rho, \ell_{\rho'}=0$, \eqref{spin2pt} reduces to \eqref{eqsc}. 

\subsection{Deformed scalar-scalar-STT 3-point functions}  
\label{STT}
For deformed three point functions with one external STT operator, we start with,
\be
\int D^d X \la\mo(X)\mo_{i }(X_1)\mo_{j}(X_2)\mo_\rho(X_3)\ra\,
\ee
where the label $\mo_\rho$ denotes an STT operator of dimension and spin $(\D_\rho,\ell_\rho)$. The deformation can hit either the scalar operators or the spin operator. When $\mo$ hits either $\mo_{i }$ or $\mo_{j}$ with the same operator exchange in the internal channel, we get the dangerous divergence of the same form as in \eqref{boo}, with the additional factor $V_{3,12}^{\ell_\rho}$ to take into account of the tensor structures. Thus,
\be\label{scal}
P_{12\rho}(X,X_1,X_2,X_3,Z_3):=C_{\mo i i }\frac{X_{12}^\frac{\D}{4}X_{13}^\frac{\D}{4}}{X_{01}^\frac{\D}{2}X_{02}^\frac{\D}{4}X_{03}^\frac{\D}{4}}\frac{C_{i j\rho}V_{3,12}^{\ell_\rho}}{X_{12}^\frac{\D_i+\D_j-\D_\rho-\ell_\rho}{2}X_{13}^\frac{\D_i+\D_\rho+\ell_\rho-\D_j}{2}X_{23}^\frac{\D_j+\D_\rho+\ell_\rho-\D_i}{2}}\,.
\ee 
and the same for $P_{21\ell_\rho}$, under the replacement $1\leftrightarrow2$ and $\D_i\leftrightarrow\D_j$. 
When the deformation hits the external STT operator, there are multiple tensor structures. However, from the deformed two point functions for two STT operators, we see that the wavefunction renormalization to remove the divergence and the $\log$ terms come with a linear combination of these tensor structures, which we call $\mc_{\rho\rho\mo}$. Thus, when the deformation hits the STT operator, we can write,
\be\label{spin}
P_{\rho 12}(X,X_1,X_2,X_3,Z_3):=C_{\rho\rho\mo}\frac{X_{23}^\frac{\D}{4}X_{13}^\frac{\D}{4}}{X_{03}^\frac{\D}{2}X_{01}^\frac{\D}{4}X_{02}^\frac{\D}{4}}\frac{C_{i j \rho}V_{3,12}^{\ell_\rho}}{X_{12}^\frac{\D_i+\D_j-\D_\rho-\ell_\rho}{2}X_{13}^\frac{\D_i+\D_\rho+\ell_\rho-\D_j}{2}X_{23}^\frac{\D_j+\D_\rho+\ell_\rho-\D_i}{2}}\,,
\ee
where we have replaced $1\leftrightarrow3$ in \eqref{scal} and swept the contribution of the entire tensor structure in the coefficient $C_{\rho\rho\mo}$. When the deformation hits the STT operator, this is effectively the contribution from the exchange of the $\mo_\rho$ primary in the internal channel. Intuitively, this is the the analog of the contribution when $\mo_\rho$ is a scalar operator at $X_3$. Following the argument of section \ref{Threept}, we can decompose the four point function in terms of the ``dangerous divergence" and the ``safe" part as,
\begin{align}
\begin{split}
\la\mo(X)\mo_{i }(X_1)\mo_{j}(X_2)\mo_\rho(X_3,Z_3)\ra=&\la\mo(X)\mo_{i }(X_1)\mo_{j}(X_2)\mo_\rho(X_3,Z_3)\ra^\circ\\
&+\sum_{(abc)=(12\rho),(21\rho),(\rho 12)} P_{abc}(X,X_1,X_2,X_3,Z_3)\,,
\end{split}
\end{align} 
where again $\la\mo(X)\mo_{i }(X_1)\mo_{j}(X_2)\mo_\rho(X_3,Z_3)\ra^\circ$ has no dangerous divergence and has correct scaling properties under $X\rightarrow cX$, $(X_1,X_2)\rightarrow (c_1X_1,c_2X_2)$ and $(X_3,Z_3)\rightarrow (c_3X_3,e Z_3)$. Hence,
\be
\int D^d X \la\mo(X)\mo_{i }\mo_{j}(X_2)\mo_\rho(X_3,Z_3)\ra^\circ=\frac{c_{i j\rho}V_{3,12}^{\ell_\rho}}{X_{12}^\frac{\D_i+\D_j-\D_\rho-\ell_\rho}{2}X_{13}^\frac{\D_i+\D_\rho+\ell_\rho-\D_j}{2}X_{23}^\frac{\D_j+\D_\rho+\ell_\rho-\D_i}{2}}\,,
\ee 
while the integral over the ``dangerous" part of the spin contribution, is of the form,
\begin{align}
\begin{split}
&\int D^d X P_{\ell_\rho 12}(X,X_1,X_2,X_3,Z_3)=\\
& \frac{\mc_{\rho\rho\mo}C_{i j\rho} V_{3,12}^{\ell_\rho}}{X_{12}^\frac{\D_i+\D_j-\D_\rho-\ell_\rho}{2}X_{13}^\frac{\D_i+\D_\rho+\ell_\rho-\D_j}{2}X_{23}^\frac{\D_j+\D_\rho+\ell_\rho-\D_i}{2}}\bigg[-\frac{1}{\e}+\frac{1}{2}\log \frac{X_{23}X_{13}}{X_{12}}+H_d+O(\e)\bigg]\,,
\end{split}
\end{align}
where $H_d$ is given in \eqref{Hfn}. Combining all the contributions, 
\be
\full{\la\mo_{i }^\text{ren}\mo_{j}^\text{ren}\mo_\rho^\text{ren}\ra=\frac{C_{i j\rho}V_{3,12}^{\ell_\rho}}{X_{12}^\frac{\D_i+\D_j-\D_\rho-\ell_\rho}{2}X_{13}^\frac{\D_i+\D_\rho+\ell_\rho-\D_j}{2}X_{23}^\frac{\D_j+\D_\rho+\ell_\rho-\D_i}{2}}}\,,
\ee
where, as in the scalar case, 
\be
\full{\mc_{i j\rho}}=\bigg[1+\lambda(\mc_{\mo ii}+\mc_{\mo jj}+\mc_{\rho\rho\mo}) H_d\bigg]C_{i j\rho}+\lambda c_{i j\rho}\,.
\ee
This implies,
 \begin{equation}
\full{\vev{\mo_{i }^\ren\mo_{j}^\ren\mo_{\rho}^\ren}=\sum_{\rho'}\fF_{i j\rho'} \vev{\mo_{\rho'}^\ren\mo_{\rho}^\ren}},\label{deformed3ptOPEspin}
\end{equation} where 
\begin{equation}
\full{\fF_{i j\rho} =C_{i j\rho} \fC }
\end{equation}
is a generalization of the deformed scalar OPE to within first order perturbation theory.

\bibliographystyle{ytphys}
\bibliography{refs}

\providecommand{\href}[2]{#2}\begingroup\raggedright\begin{thebibliography}{1}

\bibitem{Berenstein:2014cia} 
  D.~Berenstein and A.~Miller,
  ``{Conformal perturbation theory, dimensional regularization, and AdS/CFT correspondence},''
\href{http://dx.doi.org/10.1103/PhysRevD.90.086011}{\em{
  Phys.\ Rev.\ D} {\bf 90}, no. 8, 086011 (2014)}
 \href{https://arxiv.org/abs/1406.4142}{{\ttfamily arXiv:1406.4142 [hep-th]}}. 

\bibitem{Berenstein:2016avf} 
  D.~Berenstein and A.~Miller,
  ``{Logarithmic enhancements in conformal perturbation theory and their real time interpretation},''
 \href{https://arxiv.org/abs/1607.01922}{{\ttfamily  arXiv:1607.01922 [hep-th]}}.

\bibitem{Gaberdiel:2008fn}
M.~R. Gaberdiel, A.~Konechny, and C.~Schmidt-Colinet, ``{Conformal Perturbation
  Theory Beyond the Leading Order},''
  \href{http://dx.doi.org/10.1088/1751-8113/42/10/105402}{{\em J. Phys.}
  {\bfseries A42} (2009) 105402},
\href{http://arxiv.org/abs/0811.3149}{{\ttfamily arXiv:0811.3149 [hep-th]}}.

\bibitem{Bashmakov:2017rko}
V.~Bashmakov, M.~Bertolini, and H.~Raj, ``{On Non-Supersymmetric Conformal
  Manifolds: Field Theory and Holography},''
\href{http://arxiv.org/abs/1709.01749}{{\ttfamily arXiv:1709.01749 [hep-th]}}.

\bibitem{Behan:2017mwi}
C.~Behan, ``{Conformal Manifolds: ODEs from OPEs},''
\href{http://arxiv.org/abs/1709.03967}{{\ttfamily arXiv:1709.03967 [hep-th]}}.

\bibitem{Hollands:2017chb}
S.~Hollands, ``{Action Principle for OPE},''
\href{http://arxiv.org/abs/1710.05601}{{\ttfamily arXiv:1710.05601 [hep-th]}}.

\bibitem{Costa:2011mg}
M.~S. Costa, J.~Penedones, D.~Poland, and S.~Rychkov, ``{Spinning Conformal
  Correlators},'' \href{http://dx.doi.org/10.1007/JHEP11(2011)071}{{\em JHEP}
  {\bfseries 11} (2011) 071},
\href{http://arxiv.org/abs/1107.3554}{{\ttfamily arXiv:1107.3554 [hep-th]}}.

\bibitem{SimmonsDuffin:2012uy}
D.~Simmons-Duffin, ``{Projectors, Shadows, and Conformal Blocks},''
  \href{http://dx.doi.org/10.1007/JHEP04(2014)146}{{\em JHEP} {\bfseries 04}
  (2014) 146},
\href{http://arxiv.org/abs/1204.3894}{{\ttfamily arXiv:1204.3894 [hep-th]}}.



\end{thebibliography}\endgroup

\end{document}